\documentclass[10pt]{llncs}
\usepackage{epic,eepic,amsmath,latexsym,amssymb,color}
\usepackage{ifthen,graphics,epsfig,subfigure,comment}
\usepackage[english]{babel}
\usepackage{times}
\usepackage{framed}
\usepackage{array}
\usepackage{color}
\usepackage{colortbl}
\usepackage{url}
\usepackage{pgf}
\usepackage{tikz}
\usetikzlibrary{arrows,matrix, calc, fit}

\usepackage{rotating}

\usepackage{pgfplots}

\usepackage{comment}

\usepackage{enumitem}
\usepackage{amsmath}
\usetikzlibrary{patterns}
\usepackage{floatrow}
\usepackage{soul}

\begin{document}

  \def\startCirc#1{\tikz[remember picture,overlay]\path node[inner sep=0, anchor=south] (st) {#1} coordinate (start) at (st.center);}%
  \def\endCirc#1{\tikz[remember picture,overlay]\path node[inner sep=0, anchor=south] (en) {#1} coordinate (end) at (en.center);%
    \begin{tikzpicture}[overlay, remember picture]%
      \path (start);%
      \pgfgetlastxy{\startx}{\starty}%
      \path (end);%
      \pgfgetlastxy{\endx}{\endy}%
      \pgfmathsetlengthmacro{\xdiff}{\endx-\startx}%
      \pgfmathsetlengthmacro{\ydiff}{\endy-\starty}%
      \pgfmathtruncatemacro{\xdifft}{\xdiff}%
      \pgfmathsetmacro{\xdiffFixed}{ifthenelse(equal(\xdifft,0),1,\xdiff)}%
      \pgfmathsetmacro{\angle}{ifthenelse(equal(\xdiffFixed,1),90,atan(\ydiff/\xdiffFixed))}%
      \pgfmathsetlengthmacro{\xydiff}{sqrt(abs(\xdiff^2) + abs(\ydiff^2))}%
      \path node[draw,rectangle,red, rounded corners=2mm, rotate=\angle, minimum width=\xydiff+4ex, minimum height=2.5ex] at ($(start)!.5!(end)$) {};%
    \end{tikzpicture}%
  }

\newlength {\squarewidth}
\renewenvironment {square}
{
\setlength {\squarewidth} {\linewidth}
\addtolength {\squarewidth} {-12pt}
\renewcommand{\baselinestretch}{0.75} \footnotesize
\begin {center}
\begin {tabular} {|c|} \hline
\begin {minipage} {\squarewidth}
\medskip
}{
\end {minipage}
\\ \hline
\end{tabular}
\end{center}
}



\newcommand{\toto}{xxx}
\newenvironment{proofT}{\noindent{\bf
Proof }} {\hspace*{\fill}$\Box_{Theorem~\ref{\toto}}$\par\vspace{3mm}}
\newenvironment{proofL}{\noindent{\bf
Proof }} {\hspace*{\fill}$\Box_{Lemma~\ref{\toto}}$\par\vspace{3mm}}
\newenvironment{proofC}{\noindent{\bf
Proof }} {\hspace*{\fill}$\Box_{Corollary~\ref{\toto}}$\par\vspace{3mm}}

\newcounter{linecounter}
\newcommand{\linenumbering}{\ifthenelse{\value{linecounter}<10}
{(0\arabic{linecounter})}{(\arabic{linecounter})}}
\renewcommand{\line}[1]{\refstepcounter{linecounter}\label{#1}\linenumbering}
\newcommand{\resetline}[1]{\setcounter{linecounter}{0}#1}
\renewcommand{\thelinecounter}{\ifnum \value{linecounter} >
9\else 0\fi \arabic{linecounter}}

\newcommand{\SB}[1]{\noindent\textcolor{red}{{\fontfamily{phv}\selectfont SB-NOTE: #1}}}
\newcommand{\AD}[1]{\noindent\textcolor{blue}{{\fontfamily{phv}\selectfont AD-NOTE: #1}}}
\newcommand{\RL}[1]{\noindent\textcolor{blue}{{\fontfamily{phv}\selectfont RL-NOTE: #1}}}
\newcommand{\vir}[1]{``#1''}


\title{\bf Building Regular Registers with Rational Malicious Servers and Anonymous Clients -- Extended Version}
\titlerunning{Buiding Regular Registers with Malicious Servers}
\author{Antonella Del Pozzo\inst{1}, Silvia Bonomi\inst{1}%
	 , Riccardo Lazzeretti\inst{1} and Roberto Baldoni\inst{1}\inst{2}\\
}
\institute{
Research Center of Cyber Intelligence and Information Security (CIS), IT\\
Dept. of Computer and System Sciences ``Antonio Ruberti'',
Sapienza Universit\`{a} di Roma, IT\\
\texttt{\{delpozzo, bonomi, lazzeretti, baldoni\}@dis.uniroma1.it}
\and 
CINI Cybersecurity National Laboratory, Italy
}
\authorrunning{Antonella Del Pozzo et al.}
\date{}
\maketitle


\begin{abstract}
The paper addresses the problem of emulating a regular register in a synchronous distributed system where clients invoking ${\sf read}()$ and ${\sf write}()$ operations are anonymous while server processes maintaining the state of the register may be compromised by rational adversaries (i.e., a server might behave as \emph{rational malicious Byzantine} process).
We first model our problem as a Bayesian game between a client and a rational malicious server where the equilibrium depends on the decisions of the malicious server (behave correctly and not be detected by clients vs returning a wrong register value to clients with the  risk of being detected and then excluded by the computation). We prove such equilibrium exists and finally we design a protocol implementing the regular register that forces the rational malicious server to behave correctly.

\end{abstract}

\section{Introduction}
To ensure high service availability,  storage services are usually realized by replicating data at multiple locations and maintaining such data consistent. Thus, 
replicated servers represent today an attractive target for attackers that may try to compromise replicas correctness for different purposes, such as gaining access to protected data, interfering with the service provisioning (e.g. by delaying operations or by compromising the integrity of the service), reducing service availability with the final aim to damage the service provider (reducing its reputation or letting it pay for the violation of service level agreements), etc.
A compromised replica is usually modeled trough an arbitrary failure (i.e. a Byzantine failure) that is made transparent to clients by employing Byzantine Fault Tolerance (BFT) techniques.
Common approaches to BFT are based on the deployment of a sufficiently large number of replicas to tolerate an estimated number $f$ of compromised servers (i.e. BFT replication).  However, this approach has a strong limitation: a smart adversary may be able to compromise more than $f$ replicas in long executions and may get access to the entire system when the attack is sufficiently long.
To overcome this issue, Sousa et al. designed the \emph{proactive-reactive recovery} mechanism \cite{SBCNV10}. The basic idea is to periodically reconfigure the set of replicas to rejuvenate servers that may be under attack (proactive mode) and/or when a failure is detected (reactive mode).
%
%
This approach is effective in long executions but requires a fine tuning of the replication parameters (upper bound $f$ on the number of possible compromised replicas in a given period, rejuvenation window, time required by the state transfer, etc...) and the presence of secure components in the system. In addition, it is extremely costly during good periods (i.e. periods of normal execution) as a high number of replicas must be deployed independently from their real need. In other words, the system pays the cost of an attack even if the attack never takes place.

In this paper, \textit{we want to investigate the possibility to implement a distributed shared variable (i.e. a register) without making any assumption on the knowledge of the number of possible compromised replicas}, i.e. without relating the total number of replicas $n$ to the number of possible compromised ones $f$. To overcome the impossibility result of \cite{B00,MAD02-2}, we assume that (i) clients preserve their privacy and do not disclose their identifiers while interacting with server replicas (i.e. anonymous clients) and (ii) at least one server is always alive and never compromised by the attacker. 
We first model our protocol as a game between two parties, a client and a rational malicious server (i.e. a server controlled by rational adversaries) where each rational malicious server gets benefit by two conflicting goals: (i) it wants to have continuous access to the current value of the register and, (ii) it wants to compromise the validity of the register returning a fake value to a client. 
However, if the rational malicious server tries to accomplish goal (ii) it could be detected by a client and it could be excluded from the computation, precluding it to achieve its first goal. 
We prove that, under some constraints, an equilibrium exists for such game.  In addition, we design some distributed protocols implementing the register and reaching such equilibrium when rational malicious servers privilege goal (i) with respect to goal (ii). As a consequence, rational malicious servers return correct values to clients to avoid to be detected by clients and excluded by the computation and the register implementation is proved to be correct. 

The rest of the paper is organized as follows: Section \ref{sec:related} discusses related works, Section \ref{sec:systemModel} and Section \ref{sec:Problem} introduce respectively the system model and the problem statement. In Section \ref{sec:game} we model the problem as a Bayesian game and in Section \ref{sec:protocol} we provide a protocol matching the Bayesian Nash Equilibrium that works under some limited constraints, while in Section \ref{sec:other_protocols} we presents two variants of the protocol that relax the constraints , at the expense of some additional communications between the clients or protocol complexity increase. Finally, Section \ref{sec:conclusion} presents a discussion and future work.

\section{Related Work}\label{sec:related}

Building a distributed storage able to resist arbitrary failures (i.e. Byzantine) is a widely investigated research topic.
The Byzantine failure model captures the most general type of failure as no assumption is made on the behavior of faulty processes.
Traditional solutions to  build a Byzantine tolerant storage service can be divided into two categories: \emph{replicated state machines} \cite{S90} and \emph{Byzantine quorum systems} \cite{B00,MR98,MAD02,MAD02-2}. 
Both the approaches are based on the idea that the state of the storage is replicated among processes and the main difference is in the number of replicas involved simultaneously in the state maintenance protocol.
Replicated state machines approach requires that every non-faulty replica receives every request and processes requests in the same order before returning to the client \cite{S90} (i.e. it assumes that processes are able to totally order requests and execute them according to such order).
Given the upper bound on the number of failures $f$,  the replicated state machine approach requires only $2f+1$ replicas in order to provide a correct register implementation.
Otherwise, Byzantine quorum systems need just a sub-set of the replicas (i.e. \emph{quorum}) to be involved simultaneously.  The basic idea is that each operation is executed by a quorum and any two quorums must intersect (i.e. members of the quorum intersection act as witnesses for the correct execution of both the operations).
Given the number of failures $f$,  the quorum-based approach requires at least $3f+1$ replicas in order to provide a correct register implementation in a fully asynchronous system \cite{MAD02-2}.
Let us note that, in both the approaches, the knowledge of the upper bound on faulty servers $f$ is required to provide deterministic correctness guarantees.
In this paper, we follow an orthogonal approach. We are going to consider a particular case of byzantine failures and we study the cost, in terms of number of honest servers, of building a distributed storage (i.e. a register) when clients are anonymous and have no information about the number of faulty servers (i.e. they do not know the bound $f$). In particular, the byzantine processes here considered deviate from the protocol by following a strategy that brings them to optimize their own benefits (i.e., they are \emph{rational}) and such strategy has the final aim to compromise the correctness of the storage (i.e., they are \emph{malicious}).
In \cite{MSLCADW11}, the authors presented Depot, a cloud storage system able to tolerate any number of Byzantine clients or servers, at the cost of a weak consistency semantics called \emph{Fork-Join-Causal consistency} (i.e., a weak form of causal consistency).

Another different solution can rely on proactive secret sharing \cite{OY91}. Secret Sharing \cite{S79} guarantees that a secret shared by a client among $n$ parties (servers) cannot be obtained by an adversary corrupting no more than $f$ servers ($f$ imposed by the protocol). Moreover, if no more than $f$ servers are Byzantines, the client can correctly recover the secret from the shares provided by any $f+1$ servers. Recent Proactive Secret Sharing protocols, e.g. \cite{DEL16}, show that Secret sharing can be applied also to synchronous networks. Even if Proactive Secret Sharing can guarantee the privacy of the data (this is out of the scope of the paper) against up to $n-2$ passive adversaries, the solution has some limitations. 
First fo all, clients are not able to verify whether the number of Byzantines exceeds $t$ and hence understand if the message obtained is correct. 
Secondly, Secret Sharing protocols operating in a synchronous distributed system with Byzantines (active adversaries) correctly work with a small number of Byzantines and have high complexity ($f<n/2-1$ and $\mathcal{O}(n^4)$ in \cite{DEL16}). 

In \cite{AAC05}, the authors introduced the \emph{BAR (Byzantine, Altruistic, Rational) model} to represent distributed systems with heterogeneous entities like peer-to-peer networks. This model allows to distinguish between Byzantine processes (arbitrarily deviating  from the protocol, without any known strategy), altruistic processes (honestly following the protocol) and rational processes (may decide to follow or not the protocol, according to their individual utility).
Under the BAR model, several problems have been investigated (e.g. reliable broadcast \cite{CLNMAD08}, data stream gossip \cite{LCWNRAD06},  backup service through state machine replication \cite{AAC05}).
Let us note that in the BAR model the utility of a process is measured through the cost sustained to run the protocol. In particular, each step of the algorithm (especially sending messages) has a cost and the objective of any rational process is to minimize its global cost. 
As a consequence, rational \emph{selfish} processes deviate from the protocol just by skipping to send messages, if not properly encouraged by some reward. 
In contrast with the BAR model, in this paper we consider malicious rational servers that can deviate from the protocol with different objectives, benefiting from preventing the correct protocol execution rather than from saving messages. 

More recently, classical one-shot problems as leader election \cite{ADH13,AGLM14}, renaming and consensus \cite{AGLM14} have been studied under the assumption of rational agents (or rational processes). The authors provide algorithms implementing such basic building blocks, both for synchronous and asynchronous networks, under the so called \emph{solution preference} assumption i.e., agents gain if the algorithm succeeds in its execution while they have zero profit if the algorithm fails. As a consequence, processes will not deviate from the algorithm if such deviation interferes with its correctness.
Conversely, the model of rational malicious processes considered in this paper removes implicitly this assumption as they are governed by adversaries that get benefit when the algorithm fails while in \cite{ADH13,AGLM14} rational processes get benefit from the correct termination of the protocol (i.e. they are selfish according with the BAR model).

Finally, the model considered here can be seen as a particular case of BAR where rational servers take malicious actions, with the application similar to the one considered in \cite{AAC05}. However, in contrast to \cite{AAC05}, we do not assume any trusted third party to identify users, we assume that clients are anonymous (e.g., they are connected through the Tor anonymous network \cite{TOR}), and we investigate the impact of this assumption together with the rational model.
To the best of our knowledge, this is the first paper that analyzes how the anonymity can help in managing rational malicious behaviors.

\section{System Model }\label{sec:systemModel}


The distributed system is composed by a set of $n$ servers implementing a distributed shared memory abstraction and by an arbitrary large but finite set of clients $\mathcal{C}$. 
Servers are fully identified (i.e. they have associated a unique identifier $s_1, s_2 \dots s_n$) while clients are anonymous, i.e. they share the same identifier.\\
%
%
\noindent {\bf Communication model and timing assumptions.} Processes can communicate only by exchanging messages through  \emph{reliable} communication primitives, i.e. messages are not created, duplicated or dropped. 
The system is synchronous in the following sense: all the communication primitives used to exchange messages guarantee a timely delivery property.
In particular, we assume that clients communicate with servers trough a \emph{timely} reliable broadcast primitive (i.e., there exists an integer $\delta$, known by clients, such that if a client broadcasts a message $m$ at time $t$ and a server $s_i$ delivers $m$, then all the servers $s_j$ deliver $m$ by time $t + \delta$).
%
%
%
\noindent Servers-client and client-client communications are done through \vir{point-to-point} \emph{anonymous timely} channels (a particular case of the communication model presented in \cite{DFT13} for the most general case of homonyms). Considering that clients are identified by the same identifier $\ell$, when a process sends a point-to-point message $m$ to an identifier $\ell$, all the clients will deliver $m$. 
More formally, there exists an integer $\delta' \le \delta$, known by processes, such that if $s_i$ sends a message $m$ to a client identified by an identifier $\ell$ at time $t$, then all the clients identified by $\ell$ receive $m$ by time $t + \delta'$.
%
%
We assume that channels are \emph{authenticated} (\vir{oral} model), i.e. when a process identified by $j$ receives a message $m$ from a process identified by $i$, then $p_j$ knows that $m$ has been generated by a process having identifier $i$.\\
%
\noindent{\bf Failure model.} 
Servers are partitioned into two disjoint sub-sets: \emph{honest} servers and \emph{malicious} servers (\emph{attackers}). 
Honest servers behave according to the protocol executed in the distributed system (discussed in Section \ref{sec:protocol}) while malicious servers represent entities compromised by an adversary that may deviate from the protocol by dropping messages (omission failures), changing the content of a message, creating spurious messages, exchanging information outside the protocol, etc. Malicious servers are \emph{rational}, i.e. they deviate from the protocol by following a strategy that aims at increasing their own benefit (usually performing actions that may prevent the correct execution of the protocol).
We assume that rational malicious servers act independently, i.e. they do not form a coalition and each of them acts for its individual gain.
Servers may also fail by crashing and we identify as \emph{alive} the set of non crashed servers\footnote{Alive servers may be both honest or malicious.}. However, we assume that at least one honest alive server  always exists in the distributed system.

\section{Regular Registers}\label{sec:Problem}

A register is a shared variable  accessed by a set of processes, i.e. clients,  through two operations, namely ${\sf read()}$ and ${\sf write()}$. 
Informally, the ${\sf  write()}$ operation updates the value  stored in the shared variable while  the $\sf read()$ obtains the  value contained in the variable (i.e. the last written value). Every operation issued on a register is, generally, not instantaneous and it  can  be characterized  by  two events  occurring  at  its boundary:  an \emph{invocation} event and a \emph{reply} event. These events occur at two time instants (invocation  time and reply time) according  to the fictional global time.

An operation $op$  is \emph{complete} if both the  invocation event and the reply event occur (i.e. the  process executing the operation does not crash between the invocation and the reply).
Contrary, an operation $op$ is said to be \emph{failed} if it is invoked by a process that crashes before the reply event occurs. According to these time instants, it is possible to state when two operations are concurrent with respect to the real time execution.
For ease of presentation we assume the existence of a fictional global clock and the invocation time and response time of operations are defined with respect to this fictional clock.\\
Given two operations  $op$ and $op'$, and their  invocation event and reply event  times  ($t_{B}(op)$ and $t_B(op')$) and return  times ($t_E(op)$ and $t_E(op')$),  we say that $op$ \emph{precedes} $op'$ ($op \prec op'$) iff $t_E(op) < t_B(op')$. If $op$ does not precede $op'$ and $op'$  does not  precede $op$,  then $op$  and $op'$   are \emph{concurrent} ($op||op'$). 
Given a  ${\sf write}(v)$ operation,  the value $v$  is said to  be written when the operation  is complete.
 
In case of concurrency while  accessing the shared variable, the meaning of \emph{last written  value} becomes ambiguous.  Depending  on the  semantics of the operations, three types of register   have   been   defined   by  Lamport \cite{L86}:   \emph{safe}, \emph{regular} and \emph{atomic}. 
In this paper, we consider a regular register which is specified as follows:
\begin{itemize}[topsep=3pt]
\item ${\sf Termination}$:
If an alive client invokes an operation, it eventually  returns from that operation. 
\item ${\sf Validity}$:
A read operation returns the last value written before its invocation, or a value written by a write operation concurrent with it.
\end{itemize}

Interestingly,   safe, regular and atomic registers   have  the same  computational  power.  This  means that  it  is possible to implement a multi-writer/multi-reader atomic register from single-writer/single-reader  safe   registers. There are several papers in the literature discussing such transformations (e.g., \cite{CKW00,HV95,SAG94,V88,VA86} to cite a few).
In this paper, we assume that the register is single writer in the sense that no two ${\sf write}()$ operations may be executed concurrently. However, any client in the system may issue a ${\sf write}()$ operation. This is not a limiting assumption as clients may use an access token to serialize their writes\footnote{Let us recall that we are in a synchronous system and the mutual exclusion problem can be easily solved also in presence of failures.}. We will discuss in Section \ref{sec:conclusion} how this assumption can be relaxed.

\section{Modeling the Register protocol as a Game} \label{sec:game}

In a distributed system where clients are completely disjoint from servers, it is possible to abstract any register protocol as a sequence of requests made by clients (e.g. a request to get the value or a request to update the value) and responses (or replies) provided by servers, plus some local computation.
If all servers are honest, clients will always receive the expected replies and all replies will always provide the right information needed by the client to correctly terminate the protocol.
Otherwise, a compromised server can, according to its strategy, omit to send a reply or can provide bad information to prevent the client from terminating correctly. In this case, in order to guarantee a correct execution, the client tries to detect such misbehavior, react and punish the server. 
%
%
Thus, a distributed protocol implementing a register in presence of rational malicious servers can be modeled as a two-party game between a client and each of the servers maintaining a copy of the register: the client wants to correctly access  the register while the server wants to prevent the correct execution of a ${\sf read}()$ without being punished.\\
%
%
\noindent{\bf Players.} The two players are respectively the client and the server. Each player can play with a different role:
servers can be divided into \emph{honest} servers and \emph{malicious} servers while clients can be divided in those asking a \emph{risky request} (i.e., clients able to detect misbehaviors and punish server\footnote{Notice that the client ability to detect a server misbehaviors depends on the specific protocol.}) and those asking for a \emph{risk-less request} (i.e., clients unable to punish servers).\\ 
%
%
%
%
\noindent{\bf Strategies.} Players' strategies are represented by all the possible actions that a process may take.
Clients have just one strategy, identified by $\mathcal{R}$, that is \emph{request information to servers}. 
Contrarily, servers have different strategies depending on their failure state: 
\begin{itemize}[topsep=2pt]
\item malicious servers have three possible strategies: (i) $\mathcal{A}$, i.e. \emph{attack the client} by sending back wrong information (it can reply with a wrong value, with a wrong timestamp or both),  (ii) $\mathcal{NA}$, i.e. \emph{not attack the client} behaving according to the protocol and (iii) $\mathcal{S}$, i.e. \emph{be silent} omitting the answer to client's requests; 
\item honest servers have just the $\mathcal{NA}$ strategy.
\end{itemize} 

Let us note that the game between a honest client and a honest server is trivial as they have just one strategy that is to follow the protocol. Thus, in the following we are going to skip this case and we will consider only the game between a client and a rational malicious server.\\
\noindent{\bf Utility functions and extensive form of the game.} Clients and servers have opposite utility functions. In particular:
\begin{itemize}[topsep=3pt]
\item every client increases its utility when it is able to read a correct value from the register and it wants to maximize the number of successful ${\sf read}()$ operations;
\item every server increases its utility when it succeeds to prevent the client from reading a correct value, while it loses when it is detected by the client and it is punished.
\end{itemize}

\begin{figure}[t]
\center
\begin{tikzpicture}[transform shape, node distance=1.9cm, scale=0.65, font=\large]
\node at (-2.5,1.5) {$Risk-less\ request$};
\node at (-2,1.1) {$1-\theta$};
\node at (2,1.5) {$Risky\ request$};
\node at (2,1.1) {$\theta$};
\node at (-3.7,-.5) {$\mathcal{S}$}; \node at (-2.7,-.5) {$\mathcal{A}$}; \node at (-1.6,-.5) {$\mathcal{NA}$};
\node at (1.4,-.5) {$\mathcal{S}$}; \node at (2.6,-.5) {$\mathcal{A}$}; \node at (3.9,-.5) {$\mathcal{NA}$};
\node (a) at (0, 2) [fill=white] {$Client$};
\node (b) [below right of=a, fill=white] {$Server$};
\node (c) [below left of=a,fill=white] {$Server$};

\node (d) at (-5,-1.5) {$(D_c,-D_s)$};
\node (e) at (-3,-1.5) {$(-G_c,G_s)$};
\node (f) at (-1,-1.5) {$(G_c,0)$};

\node (g) at (1,-1.5) {$(D_c,-D_s)$};
\node (h) at (3,-1.5) {$(D_c,-D_s)$};
\node (i) at (5,-1.5) {$(G_c,0)$};
%
\draw [dashed](b) -- (c);
\foreach \from/\to in {a/b,a/c,e/c,d/c,f/c,g/b,h/b,i/b}
\draw [-] (\from) -- (\to);
\end{tikzpicture}
\caption{Extensive form of the game. Dashed line represents the unknown nature of requests from the risk point of view. Outcome pairs refer to client and server gains respectively.} \label{fig:extensive}
\end{figure}
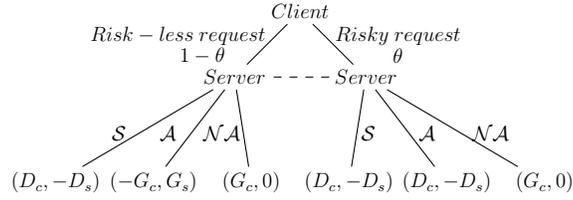

In the following, we will denote as $G_c$ the gain obtained by the client when it succeeds in reading, $G_s$ the gain obtained by the server when it succeeds in preventing the client from reading and as $D_c$ the gain of the client when detecting the server and as $D_s$ the loss of the server when it is detected. Such parameters are characteristic of every server and describe its behavior in terms of subjective gains/losses they are able to tolerate. Without loss of generality, we assume that $G_c$, $G_s$, $D_c$ and $D_s$ are all greater than $0$, that all the servers have the same $G_s$ and $D_s$\footnote{Let us note that if two servers have different values for $G_s$ and $D_s$, the analysis shown in the following is simply repeated for each server.} and that all the clients have the same $G_c$ and $D_c$.
Fig. \ref{fig:extensive} shows the extensive form of the game. 


The game we are considering is a Bayesian game \cite{FT91} as servers do not have knowledge about the client role but they can estimate the probability of receiving a risky request or a risk-less request i.e., they have a \emph{belief} about the client role.
We denote as $\theta$ (with $\theta \in [0,1]$) the server belief of receiving a risky request (i.e. the client may detect that the server is misbehaving) and with $1-\theta$ the server belief of receiving a risk-less request (i.e. the client is not be able to detect that the server is misbehaving).

\medskip


\noindent{\bf Analysis of the Bayesian Game.} In the following, we are going to analyze the existence (if any) of a \emph{Bayesian Nash Equilibrium}  i.e., a Nash Equilibrium\footnote{Let us recall that a Nash Equilibrium exists when each player selects a strategy and none of the players increases its utility by changing strategy.} computed by considering the players' belief.\\
Let us note that in our game, clients have just one strategy. Thus, the existence of the equilibrium depends only on the decisions taken by servers according to their utility parameters $G_s$, $D_s$ and their belief about the nature of a request (i.e., its evaluation of $\theta$).
%
Let us now compute the expected gain $E()$ of a server $s_i$ while selecting strategies $\mathcal{S}$, $\mathcal{NA}$ and $\mathcal{A}$:
\begin{eqnarray}
\label{eq:s}E(\mathcal{S})&=&(-D_s \times (1-\theta)) + (-D_s \times \theta) = -D_s\\ \label{eq:na}E(\mathcal{NA})&=&((1-\theta)\times0) + (\theta \times0) = 0\\
\label{eq:a}E(\mathcal{A}) &=& ((1-\theta) \times G_s) - (\theta \times D_s) 
\end{eqnarray}



\begin{lemma}\label{lem:dominatedStrat}
The strategy $\mathcal{S}$ is a dominated strategy.
\end{lemma}

\begin{proofL}
A server $s_i$ would choose to follows $\mathcal{S}$ over $\mathcal{NA}$ or $\mathcal{A}$ if (i) $E(\mathcal{S}) > E(\mathcal{A})$ or $E(\mathcal{S}) > E(\mathcal{NA})$.
However, considering that both $D_s$ and $G_s$ are grater that $0$, from equations $(1)-(3)$ we will have that $s_i$ will never choose to play $\mathcal{S}$.
\renewcommand{\toto}{lem:dominatedStrat}
\end{proofL}

It follows that servers have no gain in playing $\mathcal{S}$, whatever the other player does (cf. Lemma \ref{lem:dominatedStrat}). In fact, there would be no increment of their utility by playing $\mathcal{S}$ and then we will not consider such strategy anymore.

Let us note that a server $s_i$ would prefer to play $\mathcal{NA}$ (i.e., to behave honestly) with respect to $\mathcal{A}$ (i.e., to deviate from the protocol) when $E(\mathcal{NA})>E(\mathcal{A})$. 
Combining equations \eqref{eq:a} and \eqref{eq:na} we have that a $s_i$ would prefer to play $\mathcal{NA}$ when 

\begin{equation}\label{eq:AvsNA}
{G_s \over (G_s+D_s)} > \theta.
\end{equation}
%
%
The parameters $G_s$ and $D_s$ are strictly dependent on the attackers profile (i.e., an attacker for which is more important to stay in the system rather than subvert it or vice versa), thus we can not directly work on them. In the remaining part of the work we propose protocols to tune the $\theta$ parameter in such a way that the inequality \eqref{eq:AvsNA} holds. To this purpose, we derive the following Lemmas:

\begin{lemma}\label{lem:GmagD}
Let $s_i$ be a rational malicious server. If $D_s < G_s$ and $\theta < \frac{1}{2}$ then the best response of $s_i$ is to play strategy $\mathcal{A}$ (i.e. $\mathcal{NA}$ is a dominated strategy).
\end{lemma}

\begin{proofL}
Equation \eqref{eq:AvsNA} can be rewritten as
$$ \frac{1}{1+ \alpha} > \theta $$

where $\alpha=\frac{D_s}{G_s}$. Considering that $G_s>D_s$ it follows that $\alpha \in (0, 1)$. 
Note that $s_i$ would prefer to play $\mathcal{A}$ each time that inequality \eqref{eq:AvsNA} is satisfied and that $\theta$ is upper bounded by $\frac{1}{2}$, it follows that $s_i$ will prefer to play $\mathcal{A}$ for any $\theta \in [0, \frac{1}{2})$ and the claim follows.
\renewcommand{\toto}{lem:GmagD}
\end{proofL}

\begin{lemma}\label{lem:GminD}
Let $s_i$ be a rational malicious server. If $D_s > G_s$ and $\theta \ge \frac{1}{2}$ then the best response of $s_i$ is to never play strategy $\mathcal{A}$ (i.e. $\mathcal{NA}$ is a dominant strategy).
\end{lemma}
\begin{proofL}
Equation \eqref{eq:AvsNA} can be rewritten as
$$\frac{1}{1+ \alpha} > \theta$$

Note that $G_s<D_s$ and then $\alpha \in (1, \infty)$. 
Considering that $s_i$ would prefer to play $\mathcal{A}$ each time that inequality \eqref{eq:AvsNA} is satisfied and that $\theta$ is lower bounded by $\frac{1}{2}$, it follows that $s_i$ will never prefer to play $\mathcal{A}$ for any $\theta \in [\frac{1}{2}, 1]$ and the claim follows.

\renewcommand{\toto}{lem:GminD}
\end{proofL}

\section{A Protocol $\mathcal{P}$ for a Regular Register when $D_s \gg G_s$}\label{sec:protocol}


In this section, we propose a protocol $\mathcal{P}$ implementing a regular register in a synchronous distributed system with anonymous clients and up to $n-1$ malicious rational servers.
The protocol works under the assumption that the server loss $D_s$ in case of detection is much higher than its gain $G_s$ obtained when the client fails during a read (i.e. $D_s \gg G_s$\footnote{More precisely, $\mathcal{P}$ works when $D_s > cG_s$ where $c$ is the estimated number of clients in the system.}). This assumption models a situation where the attacker is much more interested in having access to data stored in the register and occasionally interfere with the server rather than causing a reduction of the availability (e.g., no termination or validity violation).
We will relax this assumption to the simple case $D_s > G_s$ in the next section extending $\mathcal{P}$ in two different ways.

Our protocol $\mathcal{P}$ follows the classical quorum-based approach. When a client wants to write, it sends the new value together with its timestamp to servers and waits for acknowledgments. Similarly, when it wants to read, it asks for values and corresponding timestamps and then it tries to select a value among the received ones.
Let us note that, due to the absence of knowledge on the upper bound of malicious processes, it could be impossible for a reader to select a value among those reported by servers and, in addition, the reader may be unable to distinguish well behaving servers from malicious ones.
To overcome this issue we leverage on the following observation: the last client $c_w$ writing a value $v$ is able to recognize such value while reading after its write (as long as no other updates have been performed). 
This makes the writer $c_w$ the only one able to understand which server $s_i$ is reporting a wrong value $v_i \neq v$, detect it as malicious and punish it by excluding $s_i$ from the computation. 
Thus, the basic idea behind the protocol is to exploit the synchrony of the system and the anonymity of clients to makes the writer indistinguishable from readers and \vir{force} malicious servers to behave correctly.

Let us note that anonymity itself is not enough to make the writer indistinguishable from other clients. In fact, if we consider a naive solution where we add anonymity to a register implementation (e.g., to the one given by Attiya, Bar-Noy and Dolev \cite{ABD95}), we have that servers may exploit the synchrony of the channels to estimate when the end of the write operation occurs and to infer whether a read request may arrive from the writer or from a different client (e.g., when it is received too close to a write request and before the expected end of the write). 
To this aim, we added in the ${\sf write}()$ operation implementation some \emph{dummy} read requests. These messages are actually needed to generate message patterns that make impossible to servers to distinguish messages coming from the writer from messages arriving from a different client. As a consequence, received a read request, a server $s_i$ is not able to distinguish if such request is risky (i.e. it comes from the writer) or is risk-less (i.e. it comes from a generic client).

In addition, we added a detection procedure that is executed both during ${\sf read}()$ and ${\sf write}()$ operations by any client. In particular, such procedure checks that every server answered to a request and that the reported information are \vir{coherent} with its knowledge (e.g., timestamps are not too old or too new).
The detection is done first locally, by exploiting the information that clients collect during the protocol execution, and then, when a client detects a server $s_j$, it disseminates its detection so that the malicious server is permanently removed from the computation (collaborative detection).

Finally, the timestamp used to label a new written value is updated by leveraging acknowledgments sent by servers at the end of the preceding ${\sf write}()$ operation. In particular, during each ${\sf write}()$ operation, servers must acknowledge the write of the value by sending back the corresponding timestamp. This is done on the anonymous channels that deliver such message to all the clients that will update their local timestamp accordingly.
As a consequence, any rational server is inhibited from deviating from the protocol, unless it accepts the high risk to be detected as faulty and removed from the system.

In the following, we provide a detailed description of the protocol $\mathcal{P}$ shown in Figures \ref{fig:readProtocol}-\ref{fig:detection}.\\

\begin{figure}[!t]
\centering
	\subfigure[Client Protocol] {
\fbox{
\begin{minipage}{0.4\textwidth}
\scriptsize
\resetline
\begin{tabbing}
aaaaA\=aaA\=aaaaaA\kill

{\bf Init}:~\\

\line{init-C-01} \> $replies  \leftarrow \emptyset$; $my\_last\_val \leftarrow \bot$; $my\_last\_ts \leftarrow 0$; $last\_ts \leftarrow 0$;\\ 





\line{init-C-02} \> $ack  \leftarrow \emptyset$; $honest  \leftarrow \{s_1, s_2 \dots s_n\}$; $writing \leftarrow {\sf false}$;\\ 


~~~-----------------------------------------------------------------------------------------------\\

{\bf operation}   ${\sf read}()$:~\\

\line{read-C-01} \> {\bf if} \= $(last\_ts = 0)$ \\ 

\line{read-C-02} \> \> {\bf then} \= {\bf return} $\bot$;\\

\line{read-C-03} \>\> {\bf else} \= $replies  \leftarrow \emptyset$;\\ 

\line{read-C-04} \>\>\> ${\sf broadcast}$ {\sc read()};\\

\line{read-C-05} \>\>\> {\bf wait} $(2\delta)$;\\

\line{read-C-06} \>\>\> {\bf if} \= $(\forall~ s_i \in honest, \exists <-, ts, val>~ \in replies)$\\

\line{read-C-07} \>\>\>\> {\bf then} \= ${\sf broadcast}$ {\sc readAck()};\\

\line{read-C-08} \>\>\>\>\> ${\sf return}~ val$;\\ 

\line{read-C-09} \>\>\>\> {\bf else} \= {\bf wait} $(\delta)$;\\

\line{read-C-10} \>\>\>\>\> {\bf if} \= $(\forall~ s_i \in honest, \exists <-, ts, val>~ \in replies)$\\

\line{read-C-11} \>\>\>\>\>\> {\bf then} \= ${\sf broadcast}$ {\sc readAck()};\\

\line{read-C-12} \>\>\>\>\>\>\>  ${\sf return}~ val$;\\ 

\line{read-C-13} \>\>\>\>\>\> {\bf else} \= {\bf execute}  ${\sf detection}(replies_i, R)$\\

\line{read-C-14} \>\>\>\>\>\>\>${\sf broadcast}$ {\sc readAck()};\\

\line{read-C15} \>\>\>\>\>\>\> {\bf if} \= $(\forall~ s_i \in honest, \exists <-, ts, val>~ \in replies)$\\

\line{read-C-16} \>\>\>\>\>\>\>\> {\bf then} ${\sf return}~ val$;\\ 

\line{read-C-17} \>\>\>\>\>\>\>\> {\bf else abort} ;\\

\line{read-C-18} \>\>\>\>\>\>\> {\bf endif}\\

\line{read-C-19} \>\>\>\>\> {\bf endif}\\

\line{read-C-20} \>\>\> {\bf endif}\\

\line{read-C-21} \> {\bf endif}\\

~~~------------------------------------------------------------------------------------------------\\
 
{\bf when} \= {\sc reply$(<j,ts, v, ots, ov>)$} \= is  ${\sf delivered}$:\\

\line{read-C-22} \> $replies \leftarrow replies \cup \{<j, ts, v>\}$;\\ 

\line{read-C-23} \> $replies \leftarrow replies \cup \{<j, ots, ov>\}$;\\

~~~------------------------------------------------------------------------------------------------\\
 
 {\bf when} \= {\sc detected$(s_j)$} \= is  ${\sf delivered}$:\\ 
 
 \line{C-01} $honest  \leftarrow honest \setminus \{s_j\}$;

\end{tabbing}
\normalsize
\end{minipage}

}
\label{sfig:cread}
}
 \subfigure[Server Protocol]{
\fbox{

\begin{minipage}{0.4\textwidth}
\scriptsize
\resetline
\begin{tabbing}
aaaaA\=aaA\=aaaaaA\kill


{\bf Init}:~~~~~~~~~~~~~~~~~~~~~~~~~~~~~~~~~~~~~~~~~~~~~~~~~~~~~~~~~~~~~~~~~~~~~~~~~~~~~~~~~~~~~~~~~~~~~~~~~~~~~~~~~~~~~~~~~~~~~~~~~\\

\line{init-S-01} \> $val_i \leftarrow \emptyset$; $ts_i \leftarrow 0$;\\


\line{init-S-02} \> $old\_val_i \leftarrow \bot$;  $old\_ts_i \leftarrow 0$; $reading_i \leftarrow 0$;\\



~~~------------------------------------------------------------------------------------------------\\

{\bf when} \= {\sc read$()$} \= is  ${\sf delivered}$:\\

\line{read-S-01} \> $reading_i \leftarrow reading_i+1$;\\

\line{read-S-02} \> ${\sf send}$ {\sc reply} $(<i, ts_i, val_i, old\_ts_i, old\_val_i>)$;\\

~~~------------------------------------------------------------------------------------------------\\

{\bf when} \= {\sc readAck$()$} \= is  ${\sf delivered}$:\\

\line{read-S-03} \> $reading_i \leftarrow reading_i-1$;

\end{tabbing}
\normalsize
\end{minipage}

}
\label{sfig:sread}
}
\caption{The ${\sf read}()$ protocol for a synchronous system.}
\label{fig:readProtocol}   
\end{figure}

\noindent {\bf The ${\sf read}()$ operation (Fig. \ref{fig:readProtocol}).} 
When a client wants to read, it first checks if the $last\_ts$ variable is still equal to $0$. If so, then there is no ${\sf write}()$ operation terminated before the invocation of the ${\sf read}()$ and the client returns the default value $\bot$ (line \ref{read-C-02}, Fig. \ref{sfig:cread}).
Otherwise, $c_i$ queries the servers to get the last value of the register by sending a {\sc read}$()$ message (line \ref{read-C-04}, Fig. \ref{sfig:cread}) and remains waiting for $2\delta$ times, i.e. the maximum round trip message delay (line \ref{read-C-05}, Fig. \ref{sfig:cread}).\\
%
%
When a server $s_i$ delivers a {\sc read}$()$ message, the $reading_i$ counter is increased by one and then $s_i$ sends a {\sc reply}$(<i, ts_i, val_i, old\_ts_i, old\_val_i>)$ message containing the current and old values and timestamp stored locally (lines \ref{read-S-01} - \ref{read-S-02}, Fig. \ref{sfig:sread}). \\
When the reading client delivers a {\sc reply}$(<j, ts, val, ots, ov>)$ message, it stores locally the reply in two tuples containing respectively the current and the old triples with server id, timestamp and corresponding value (lines \ref{read-C-22} - \ref{read-C-23}, Fig. \ref{sfig:cread}).
When the reader client is unblocked from the wait statement, it checks if there exists a pair $<ts, val>$ in the $replies$ set that has been reported by all servers it believes honest (line \ref{read-C-06}, Fig. \ref{sfig:cread}) and, in this case, it sends a {\sc read\_ack}$()$ message (line \ref{read-C-07}, Fig. \ref{sfig:cread}) and it returns the corresponding value (line \ref{read-C-08}, Fig. \ref{sfig:cread}).
Received the {\sc read\_ack}$()$ message, a server $s_i$ just decreases by one its $reading_i$ counter (line \ref{read-S-03}, Fig. \ref{sfig:sread}).
Otherwise, a ${\sf write}()$ operation may be in progress. To check if it is the case, the client keeps waiting for other $\delta$ time units and then checks again if a good value exists (lines \ref{read-C-09} - \ref{read-C-10}, Fig. \ref{sfig:cread}).
If, after this period, the value is not yet found, it means that some of the servers behaved maliciously. Therefore, the client executes the ${\sf detection}()$ procedure to understand who is misbehaving (cfr. Fig. \ref{fig:detection}).
Let us note that such procedure cleans up the set of honest servers when they are detected to be malicious. Therefore, after the execution of the procedure, the reader checks for the last time if a good value exists in its $replies$ set and, if so, it returns such value (line \ref{read-C-16}, Fig. \ref{sfig:cread}); otherwise the special value ${\sf abort}$ is returned (line \ref{read-C-17}, Fig. \ref{sfig:cread}).
In any case, a {\sc read\_ack}$()$ is sent to block the forwarding of new values at the server side (line \ref{read-C-14}, Fig. \ref{sfig:cread}).
%
%
%

\begin{figure}[!t]
\centering
	\subfigure[Client Protocol]{
\fbox{
\begin{minipage}{0.4\textwidth}
\scriptsize
\resetline
\begin{tabbing}
aaaA\=aA\=aA\=aaaA\kill

{\bf operation} ${\sf write}(v)$: ~~ \\


\line{write-C-01} \> $writing \leftarrow {\sf true}$; $ack \leftarrow \emptyset$;\\

\line{write-C-02} \> $my\_last\_ts \leftarrow last\_ts+1$; $my\_last\_val \leftarrow v$;\\


\line{write-C-03} \> ${\sf broadcast}$ {\sc write($<my\_last\_val, my\_last\_ts>$)};\\

\line{write-C-04} \> ${\sf wait} (\delta)$;\\

\line{write-C-05} \> $replies \leftarrow \emptyset$;\\

\line{write-C-06} \> ${\sf broadcast}$ {\sc read()};\\

\line{write-C-07} \> ${\sf wait} (\delta)$;\\

\line{write-C-08} \> ${\sf broadcast}$ {\sc read()};\\

%
%

\line{write-C-09} \> {\bf execute} ${\sf detection}(ack, A)$;\\

\line{write-C-10} \> ${\sf wait} (\delta)$;\\

\line{write-C-11} \> {\bf execute}  ${\sf detection}(replies_i, R)$;\\

\line{write-C-12} \>${\sf broadcast}$ {\sc readAck()};\\

\line{write-C-13} \>${\sf broadcast}$ {\sc readAck()};\\

\line{write-C-14} \> $writing \leftarrow {\sf false}$;\\

\line{write-C-15} \> ${\sf return} (ok)$.\\

~~~-------------------------------------------------------------------------------------------------------------------------\\
{\bf when} \= {\sc write\_ack}$(ts, s_j)$ is  ${\sf delivered}$: \\
%

\line{write-C-16} \> {\bf if} \= $(ts \ge my\_last\_ts)$ {\bf then} \= $ack \leftarrow ack \cup \{<j, ts, ->\}$ {\bf endif}\\


~~~-------------------------------------------------------------------------------------------------------------------------\\

{\bf when} \= $\exists ~ ts$ {\bf such that} $S=\{j | \exists <j, ts', -> \in ack\} \wedge S\supseteq honest$: \\


\line{write-C-17} \> {\bf if} \= $(ts \ge last\_ts)$ {\bf then} \= $last\_ts \leftarrow ts$ {\bf endif}\\




\line{write-C-18} \> {\bf for each} \= $<j, ts', -> ~\in ack$ {\bf such that} $ts'=ts$ {\bf do} $ack \leftarrow ack \setminus <j, ts', -> $ {\bf endFor}.



\end{tabbing}
\normalsize
\end{minipage}

}
\label{sfig:cwrite}
}
 \subfigure[Server Protocol]{
\fbox{

\begin{minipage}{0.4\textwidth}
\scriptsize
\resetline
\begin{tabbing}
aaaA\=aA\=aA\=aaA\kill


{\bf when} \= {\sc write$(<val, ts>)$} is  ${\sf delivered}$: ~~~~~~~~~~~~~~~~~~~~~~~~~~~~~~~~~~~~~~~~~~~~~~~~~~~~~~~~~~~~~~~~~~~~~~~~~~~~~~~~~~~~~~~~~~~~~~~~~\\

\line{write-S-01} \>   {\bf if} \= $(ts > ts_i)$ \\

\line{write-S-02} \>\>  {\bf then} \= $old\_ts_i \leftarrow ts_i$; \\

\line{write-S-03} \>\>\> $old\_val_i \leftarrow val_i$; \\

\line{write-S-04} \>\>\> $ts_i \leftarrow ts$; \\

\line{write-S-05} \>\>\> $val_i \leftarrow \{val\}$; \\


\line{write-S-06} \>\>  {\bf else} \= {\bf if} \= $(ts_i=ts)$ {\bf then} $val_i \leftarrow val_i \cup \{val\}$; {\bf endif}\\



\line{write-S-07} \>{\bf endIf}\\

\line{write-S-08} \> ${\sf send}$ {\sc write\_ack}$(ts, i)$;\\


\line{write-S-09} \> {\bf if} \= $(reading_i> 0)$ {\bf then} ${\sf send}$ {\sc reply} $(<i, ts_i, val_i, old\_ts_i, old\_val_i>)$ {\bf endif}.



\end{tabbing}
\normalsize
\end{minipage}

}
\label{sfig:swrite}
}
\caption{${\sf write}()$ protocol for a synchronous system.}
\label{fig:writeProtocol}   
\end{figure}


\noindent {\bf The ${\sf write}()$ operation (Fig. \ref{fig:writeProtocol}).} 
When a client wants to write, it first sets its $writing$ flag to ${\sf true}$, stores locally the value and the corresponding timestamp, obtained incrementing by one the current timestamp stored in $last\_ts$ variable (lines \ref{write-C-01} - \ref{write-C-02}, Fig. \ref{sfig:cwrite}), sends a {\sc write}$()$ message to servers, containing the value to be written and the corresponding timestamp (line \ref{write-C-03}, Fig. \ref{sfig:cwrite}), and remains waiting for $\delta$ time units.\\
When a server $s_i$ delivers a {\sc write}$(v, ts)$ message, it checks if the received timestamp is greater than the one stored in the $ts_i$ variable.
If so, $s_i$ updates its local variables keeping the current value and timestamp as old and storing the received ones as current (lines \ref{write-S-02} - \ref{write-S-05}, Fig. \ref{sfig:swrite}).
Contrarily, $s_i$ checks if the timestamp is the same stored locally in $ts_i$. If this happens, it just adds the new value to the set $val_i$ (line \ref{write-C-06}, Fig. \ref{sfig:swrite}).
In any case, $s_i$ sends back an {\sc ack}$()$ message with the received timestamp (lines \ref{write-S-08}, Fig. \ref{sfig:swrite}) and forwards the new value if some ${\sf read}()$ operation is in progress (lines \ref{write-S-09}, Fig. \ref{sfig:swrite}).
Delivering an {\sc ack}$()$ message, the writer client checks if the timestamp is greater equal than its $my\_last\_ts$ and, if so, it adds a tuple $<j, ts, ->$ to its $ack$ set (line \ref{write-C-16}, Fig. \ref{sfig:cwrite}).\\
When the writer is unblocked from the wait statement, it sends a {\sc read}$()$ message, waits for $\delta$ time units and sends another {\sc read}$()$ message (lines \ref{write-C-06} - \ref{write-C-08}, Fig. \ref{sfig:cwrite}). This message has two main objectives: (i) create a message pattern that makes impossible to malicious servers to distinguish a real reader from the writer and (ii) collect values to detect misbehaving servers. In this way, a rational malicious server, that aims at remaining in the system, is inhibited from misbehaving as it could be detected from the writer and removed from the computation. 
The writer, in fact, executes the ${\sf detection}()$ procedure both on the  $ack$ set and on the $replies$ set collected during the ${\sf write}()$ (lines \ref{write-C-09} - \ref{write-C-11}, Fig. \ref{sfig:cwrite}).
Finally, the writer sends two {\sc read\_ack}$()$ messages to block the forwarding of replies, resets its $writing$ flag to ${\sf false}$ and returns from the operation (lines \ref{write-C-12} - \ref{write-C-15}, Fig. \ref{sfig:cwrite}).\\
Let us note that, the execution of a ${\sf write}()$ operation triggers the update of the $last\_ts$ variable at any client. This happens when in the $ack$ set there exists a timestamp reported by any honest server (lines \ref{write-C-17} - \ref{write-C-18}, Fig. \ref{sfig:cwrite}).\\

\begin{figure}[!t]
\centering

\fbox{
\begin{minipage}{0.4\textwidth}
\scriptsize
\resetline
\begin{tabbing}
aaaA\=aA\=aA\=aaaA\kill

{\bf procedure} ${\sf detection}(replies\_set, set\_type)$: ~~ \\

\line{detection-01} \> $S=\{j | \exists <j, -, -> \in replies\_set \}$;\\

\line{detection-02} \> {\bf if} \= $(honest \not \subseteq S)$\\

\line{detection-03} \>\> {\bf then} \= {\bf for each} \= $s_j \in (honest_i \setminus S)$ {\bf do}\\

\line{detection-04} \>\>\>\> {\bf trigger} ${detect}(s_j)$;\\

\line{detection-05} \>\>\>\> $honest_i \leftarrow honest_i \setminus \{s_j\}$;\\

\line{detection-06} \>\>\>\> ${\sf broadcast}$ {\sc detected}$(s_j)$;\\

\line{detection-07} \>\>\> {\bf endFor}\\

\line{detection-08} \> {\bf endif}\\

\line{detection-09} \> {\bf if} \=$(set\_type = R)$  \\

\line{detection-10} \>\> {\bf then} \= {\bf if} \= ($writing$)\\


\line{detection-11} \>\>\>\> {\bf then} \= $R=\{j | \exists <j, my\_last\_val, my\_last\_ts> \in replies\_set \}$;\\

\line{detection-12} \>\>\>\>\> {\bf if} \= $(honest \not \subseteq R)$\\

\line{detection-13} \>\>\>\>\>\> {\bf then} \= {\bf for} \= {\bf each}  $s_j \in (honest_i \setminus R)$ {\bf do}\\

\line{detection-14} \>\>\>\>\>\>\>\> {\bf trigger} ${detect}(s_j)$;\\

\line{detection-15} \>\>\>\>\>\>\>\> $honest_i \leftarrow honest_i \setminus \{s_j\}$;\\

\line{detection-16} \>\>\>\>\>\>\>\> ${\sf broadcast}$ {\sc detected}$(s_j)$;\\

\line{detection-17} \>\>\>\>\>\>\> {\bf endFor}\\

\line{detection-18} \>\>\>\>\> {\bf endIf}\\


\line{detection-19} \>\>\>\> {\bf else} \= {\bf for each} \= $<j, ts, -> \in replies\_set$ {\bf such that} $ts < last\_ts-1$ {\bf do}\\

\line{detection-20} \>\>\>\>\>\> {\bf trigger} ${detect}(s_j)$;\\

\line{detection-21}\>\>\>\>\>\> $honest \leftarrow honest \setminus \{s_j\}$;\\

\line{detection-22} \>\>\>\>\>\> ${\sf broadcast}$ {\sc detected}$(s_j)$;\\

\line{detection-23} \>\>\>\>\> {\bf endFor}\\

\line{detection-24} \>\>\>\>\> {\bf for each} \= $<j, ts, val> \in replies\_set$ {\bf such that} $ts =my\_last\_ts$ {\bf do}\\

\line{detection-25} \>\>\>\>\>\> $D_i=\{v  ~|~ (\exists <j, ts, val>\in replies\_set) \wedge (ts = my\_last\_ts)\}$;\\

\line{detection-26} \>\>\>\>\>\> {\bf if} \= $((my\_last\_val \neq \bot) \wedge (my\_last\_ts=last\_ts) \wedge (last\_val \notin D_i))$\\

\line{detection-27} \>\> \>\>\>\>\> {\bf then} \= {\bf trigger} ${detect}(s_j)$;\\

\line{detection-28} \>\>\>\>\>\>\>\> $honest \leftarrow honest \setminus \{s_j\}$;\\

\line{detection-29} \>\>\>\>\>\>\>\> ${\sf broadcast}$ {\sc detected}$(s_j)$;\\

\line{detection-30} \>\>\>\>\>\> {\bf endif}\\

\line{detection-31} \>\>\>\>\> {\bf endFor}\\

\line{detection-32} \>\>\>\>\> {\bf for each} \= $<j, ts, val> \in replies\_set$ {\bf such that} $ts > last\_ts+1$ {\bf do}\\

\line{detection-33} \>\>\>\>\>\>  {\bf trigger} ${detect}(s_j)$;\\

\line{detection-34} \>\>\>\>\>\> $honest_i \leftarrow honest_i \setminus \{s_j\}$;\\

\line{detection-35} \>\>\>\>\>\> ${\sf broadcast}$ {\sc detected}$(s_j)$;\\

\line{detection-36} \>\>\>\>\> {\bf endFor}\\

\line{detection-37} \>\>\> {\bf endif}\\

\line{detection-38} \>\> {\bf else} \= {\bf for each} \= $<j, ts, -> \in replies\_set$ {\bf such that} $ts \neq my\_last\_ts$ {\bf do}\\

\line{detection-39} \>\>\>\> {\bf trigger} ${detect}(s_j)$;\\

\line{detection-40} \>\>\>\> $honest \leftarrow honest \setminus \{s_j\}$;\\

\line{detection-41} \>\>\>\> ${\sf broadcast}$ {\sc detected}$(s_j)$;\\

\line{detection-42} \>\>\> {\bf endFor}\\

\line{detection-43} \> {\bf endif}.

\end{tabbing}

\end{minipage}

}

\caption{${\sf detection}()$ function invoked by an anonymous client for a synchronous system.}
\label{fig:detection}   
\end{figure}
 

\noindent{\bf The ${\sf detection}()$ procedure (Fig \ref{fig:detection}).} This procedure is used by clients to detect servers misbehaviors during the execution of ${\sf read}()$ and ${\sf write}()$ operations.
It takes as parameter a set (that can be the $replies$ set or the $ack$ set) and a flag that identifies the type of the set (i.e. $A$ for ack, $R$ for replies).
%
In both cases, the client checks if it has received at least one message from any server it saw honest and detects as faulty all the servers omitting a message (lines \ref{detection-01} - \ref{detection-08}).\\
If the set to be checked is a set of {\sc ack}$()$ messages, the client (writer) just checks if some server $s_j$ acknowledged a timestamp that is different from the one it is using in the current ${\sf write}()$ and, if so, $s_j$ is detected as malicious (lines \ref{detection-38} - \ref{detection-42}).
Otherwise, if the set is the $replies$ set (flagged as $R$), the client checks if it is running the procedure while it is writing or reading (line \ref{detection-10}).
If the client is writing, it just updated the state of the register. Thus, the writer checks that all servers sent back the pair $<v, ts>$ corresponding to the one stored locally in the variables $my\_last\_val$ and $my\_last\_ts$. If someone reported a bad value or timestamp, it is detected as misbehaving (lines \ref{detection-11} - \ref{detection-18}).
If the client is reading, it is able to detect servers sending back timestamps that are too old (lines \ref{detection-19} - \ref{detection-23}) or too new to be correct (lines \ref{detection-32} - \ref{detection-36}) or servers sending back the right timestamp but with a wrong value (lines \ref{detection-24} - \ref{detection-31}).


\subsection{Correctness Proofs for $\mathcal{P}$}
In the following we prove that the protocol presented in Fig. \ref{fig:readProtocol} - \ref{fig:detection} terminates (Lemma \ref{lem:writeTerm}, Lemma \ref{lem:readTerm} and Theorem \ref{th:termination}), and demonstrate some properties of the timestamp mechanism used to label ${\sf write}()$ operations. In particular, we  prove that the protocol ensures the increasing monotonic order of timestamps (Lemma \ref{lem:monotonicTS}) and the consistency of the variable storing the last used timestamp (Lemma \ref{lem:writeVal}).
In Lemma \ref{lem:effectiveWrite} we prove that the last written value persists locally at each honest server, while Theorem \ref{t:corretti} proves that if servers behave honestly then the protocol emulates a regular register.
Then we prove the accuracy of the detection function (Lemma \ref{lemm:d2}, Lemma \ref{lemm:d3} and Theorem \ref{t:detection}) and that the proposed protocol $\mathcal{P}$ emulates a regular register under specific conditions (Theorem \ref{th:final}). 

\begin{lemma}\label{lem:writeTerm}
Let $c_{\ell}$ be an anonymous client invoking a ${\sf write}()$ operation. If $c_{\ell}$ executes the protocol in Fig. \ref{fig:writeProtocol} then it eventually returns from the ${\sf write}()$ operation.
\end{lemma}

\begin{proofL}
The proof simply follows by observing that in the ${\sf write}()$ operation code (e.g. Fig. \ref{sfig:cwrite}) the return event happens after three ${\sf wait}()$ statement. Thus, considering that $c_{\ell}$ is honest, it will be unblocked, from the last ${\sf wait}()$ statement, $3\delta$ time units after the ${\sf write}()$ invocation and the claim follows.
\renewcommand{\toto}{lem:writeTerm}
\end{proofL}

\begin{lemma}\label{lem:readTerm}
Let $c_{\ell}$ be an anonymous client invoking a ${\sf read}()$ operation. If $c_{\ell}$ executes the protocol in Fig. \ref{fig:readProtocol} then it eventually returns from the ${\sf read}()$ operation.
\end{lemma}

\begin{proofL}
The proof simply follows by observing that in the ${\sf read}()$ operation code a return event is defined in every branch of the code and it only
 depends on ${\sf wait}()$ statements. Thus, considering that $c_{\ell}$ is honest, it will be unblocked in a finite time after the ${\sf read}()$ invocation and the claim follows.
\renewcommand{\toto}{lem:readTerm}
\end{proofL}

\begin{theorem}[Termination]\label{th:termination}
Let $c_{\ell}$ be an anonymous client invoking an operation $op$. If $c_{\ell}$ executes the protocol in Fig. \ref{fig:readProtocol} - \ref{fig:detection} then it eventually returns from $op$.
\end{theorem}

\begin{proofT}
The proof directly follows from Lemma \ref{lem:writeTerm} and Lemma \ref{lem:readTerm}.
\renewcommand{\toto}{th:termination}
\end{proofT}

\begin{lemma}\label{lem:ackDetection}
Let $op$ be a ${\sf write}()$ operation and let $ts$ be the timestamp associated by the writer to $op$.
If a malicious server $s_i$ deviates from the protocol by omitting the {\sc write\_ack}$(ts)$ message or by sending a {\sc write\_ack}$(ts')$ message (with $ts'\neq ts$), then $s_i$ will be detected as malicious by any client.
\end{lemma}

\begin{proofL}
The ${\sf detection}()$ function is executed by the writer client on the $ack$ set at line \ref{write-C-09}, Fig. \ref{sfig:cwrite}.
The $ack$ set is emptied at the beginning of every ${\sf write}()$ operation and it is filled-in by the writer when it deliver {\sc write\_ack}$(ts')$ messages.
Such messages are sent by servers when delivering a {\sc write}$(<ts, val>)$ message sent by the writer at the beginning of the operation.
The proof simply follows by considering that the writer client knows the real value of the timestamp associated to the write and stored in its $my\_last\_ts$ local variable (line \ref{write-C-02}, Fig. \ref{sfig:cwrite}).
Thus, when the writer executes the ${\sf detection}()$ function the writer checks (i) if it has received a {\sc write\_ack}$()$ message from any servers it sees (line \ref{detection-01}-\ref{detection-07}, Fig. \ref{fig:detection})  and (ii) if all the alive servers acknowledge the right timestamp (line \ref{detection-38}-\ref{detection-42}, Fig. \ref{fig:detection}).
Considering that the communications are timely and the detection happens $2\delta$ time units after the broadcast of the {\sc write}$(<ts, val>)$ message (i.e. after the maximum round trip delay), if some servers do not answer they are detected as malicious as they omitted to answer.
In the second case, since the writer know its timestamp and channel are authenticated, it is able to detect as malicious the server answering with a different timestamp.
Finally, considering that (i) the writer notifies to all the other clients its detections, (ii) such detections are done $\delta$ time units before the end of the write and (iii) clients notifications delay is also bounded by $\delta$, it follows that at the end of the write any client detected the malicious servers and the claim follows.
\renewcommand{\toto}{lem:ackDetection}
\end{proofL}

\begin{lemma}\label{lem:writeVal}
At the end of every ${\sf write}()$ operation any client stores in its $last\_ts$ variable the same timestamp.
\end{lemma}


\begin{proofL}
Every client initializes its $last\_ts$ variable to $0$ during the init phase (line \ref{init-C-01}, Fig. \ref{sfig:cread}).
Such variable is updated at line \ref{write-C-17}, Fig. \ref{sfig:cwrite} when the client stores in its $ack$ set a timestamp $ts'$ that has been acknowledged by any alive server and that is greater than the previous one (to preserve the monotonically increasing order of timestamps). 
Then, we just need to prove that if a client updates its $last\_ts$ variable then all the clients will update it as well.

Considering that (i) {\sc write\_ack}$(ts')$ messages are sent by servers when a {\sc write}$(<val, ts'>)$ message is delivered,  (ii) {\sc write}$(<val, ts'>)$ messages are sent to all the servers and (iii) {\sc write\_ack}$(ts')$ messages are sent through point-to-point anonymous channels, we have that all clients receive {\sc write\_ack}$(ts')$  messages from the same set of servers.
In addition, due to Lemma \ref{lem:ackDetection}, we have that the set of alive processes is shared by every client and the claim follows.

\renewcommand{\toto}{lem:writeVal}
\end{proofL}

\begin{lemma}\label{lem:monotonicTS}
Let $op$ and $op'$ be two ${\sf write}()$ operations such that $op \prec op'$. Let $ts$ and $ts'$ be respectively the timestamp associated to $op$ and to $op'$, then $ts < ts'$.
\end{lemma}

\begin{proofL}
The proof simply follows by Lemma \ref{lem:writeVal} and considering that the timestamp associated to a ${\sf write}()$ operation is computed by incrementing the $last\_ts$ variable by one.
\renewcommand{\toto}{lem:monotonicTS}
\end{proofL}

\begin{corollary}\label{cor:singWriterNoCrash}
If there not exists two concurrent ${\sf write}()$ operations, for any pair of ${\sf write}()$ $w_1$, $w_2$ such that $w_1 \prec w_2$ and there not exists any $w_3$ such that $w_1 \prec w_3 \prec w_2$ then the timestamp $ts_2$ associated to $w_1$ is equal to $ts_1+1$.
\end{corollary}

\begin{proofC}
The proof simply follows by considering that timestamps are computed by incrementing the $last\_ts$ variable and it is updated at most once during each ${\sf write}()$ operation.
\renewcommand{\toto}{cor:singWriterNoCrash}
\end{proofC}

\begin{lemma}\label{lem:effectiveWrite}
At the end of a ${\sf write}(v)$ operation, every server $s_i$ behaving honestly stores the value $v$ in its $val_i$ local variable.
\end{lemma}

\begin{proofL}
Every server $s_i$ updates its $value_i$ variable in line \ref{write-S-05} or line \ref{write-S-06}, Fig. \ref{sfig:swrite}. In particular, this happens when the timestamp attached to the {\sc write}$()$ message and associated to the ${\sf write}()$ operation is greater equal than the one stored in $ts_i$.
Due to Lemma \ref{lem:monotonicTS}, timestamps follows a monotonically increasing order and thus every ${\sf write}()$ operation will have a timestamp that is greater or equal than the one previously stored.
As a consequence, when delivering a {\sc write}$(<val, ts'>)$ message, any server $s_i$ will always execute line \ref{write-S-05} or \ref{write-S-06}, Fig. \ref{sfig:swrite} storing locally the new value and the claim follows.
\renewcommand{\toto}{lem:effectiveWrite}
\end{proofL}

To the ease of presentation, let us assume that the default value $\bot$ is written by a fictional instantaneous ${\sf write}(\bot)$ operation preceding every operation $op$ executed by clients and let us define a valid value as follows:

\begin{definition}
Let $v$ be the value returned by a ${\sf read}()$ operation $op$ invoked on the regular register. $v$ is said to be \emph{valid} if\\ 
\indent (i) it is the value written by the last ${\sf write}()$ operation terminated before $op$ or\\ 
\indent (ii) it is the value written by a ${\sf write}()$ operation concurrent with $op$.
\end{definition}

\begin{lemma}\label{lem:validityCorr}
If all the servers behave honestly (i.e. they follow the protocol presented in Fig. \ref{sfig:sread} - \ref{sfig:swrite}) then any ${\sf read}()$ operation returns a valid value.
\end{lemma}

\begin{proofL}
Let us suppose by contradiction that all servers follow the protocol and that there exists a ${\sf read}()$ operation $op$ that returns a value $v$ that is not valid.\\
Let $v\_old$ be the value written by the last ${\sf write}()$ terminated before the invocation of $op$.

If $v$ is not valid, it means that $v$ is different from $v\_old$ and from the value $v'$ written by a concurrent ${\sf write}(v')$, if it exists.\\

\noindent {\bf Case 1 - No ${\sf write}()$ operation is concurrent with $op$.} Due to Lemma \ref{lem:effectiveWrite}, at time $t$ when $op$ is invoked, any server will store locally in their $value_i$ variable the value $v\_old$, together with its timestamp.
Since the value stored in the register is updated only when a {\sc write}$()$ message is delivered (line \ref{write-S-05} or \ref{write-S-06}, Fig. \ref{sfig:swrite}), and this happens only when a ${\sf write}()$ operation is triggered (line \ref{write-C-03}, Fig. \ref{sfig:cwrite}), we have that, if no ${\sf write}()$ operation is concurrent with $op$, $v\_old$, together with its timestamp, will be stored and not updated by any server during the whole execution of $op$.
Considering that any server $s_i$ behaves honestly, while delivering a {\sc read}$()$ message, $s_i$ will answer by sending a {\sc reply}$(<i, ts_i, value_i>)$ containing the same value and the same timestamp to the reader (line \ref{read-S-02}, Fig. \ref{sfig:sread}).
Thus, at time $t+2 \delta$ the $replies$ set will contains $n$ tuple $< v\_old, ts >$ and the condition at line \ref{read-C-06}, Fig. \ref{sfig:cread} holds terminating the ${\sf read}()$ operation with $v\_old$ and we have a contradiction as it is a valid value.\\

\noindent {\bf Case 2 - There exists a ${\sf write}(v')$ operation $op'$ concurrent with $op$.}
Without loss of generality, let $x$ be the timestamp associated to the value $v\_old$ written by the last terminated ${\sf write}()$ operation preceding $op$.
Let us denote by $op'$ the ${\sf write}(v')$ operation concurrent with the ${\sf read}()$ operation $op$.
Let us note that, according to the protocol in Fig. \ref{sfig:cread}, while executing the ${\sf read}()$ operation $op$, the reader client will inquiry servers to get the value of the register together with its timestamp.
Considering that, by assumption, any alive server $s_i$ behaves honestly, it follows that delivering a {\sc read}$()$ message $s_i$ will answer by sending back a {\sc reply}$()$ message containing both the current and the old value and timestamp.
Such values are modified only al line \ref{write-S-05} or \ref{write-S-06}, Fig. \ref{sfig:swrite} when a server $s_i$ deliver a {\sc write}$()$ message.

Let $t_B(op')$ and $t_B(op)$ the time at which respectively $op'$  and $op$ are invoked and let $t_E(op')$ and $t_E(op)$ be respectively the return time of $op'$ and $op$.

Let us consider the following cases:

\begin{itemize}
\item {\bf Case 2.1 -  $\mathbf{t_B(op')+\delta < t_B(op) <  t_B(op') +2\delta}$.} \\
At the beginning of the ${\sf write}()$ (i.e. at time $t_B(op')$), the writer client sends a {\sc write}$()$ message (line \ref{write-C-03}, Fig. \ref{sfig:cwrite}) that will be delivered by any alive server by time $t_B(op')+\delta$. 
When a server $s_i$ receives a {\sc write}$()$ message, it will update its $val_i$ variable to $v'$ and its $ts_i$ to the current timestamp\footnote{Let us note that  such variable will be always updated, as in line \ref{write-S-05} or \ref{write-S-06}, Fig. \ref{sfig:swrite}, due to Lemma \ref{lem:monotonicTS}.}.
It follow that from time $t_B(op')+\delta$ any alive server will store locally the value $v'$ written by $op'$. Let us note that, since $op'$ lasts until time $t_E(op')= t_B(op')+3\delta$ and, by assumption, there not exist concurrent ${\sf write}()$ operations, such variables will not be updated anymore before time $t_B(op')+3\delta$.

When an alive server $s_i$ delivers a {\sc read}$()$ message (between time $t_B(op')+\delta$ and $t_B(op')+2\delta$), it executes line \ref{read-S-02}, Fig. \ref{sfig:sread} sending to the reader both the current and the old values and timestamp through a {\sc reply}$()$ message.
Delivering such {\sc reply}$()$ message, the reader will store the values in its $replies$ local variable and waits until time $t_B(op)+2\delta$ before checking the content of such variable.
Considering that, by assumption, all the servers behave honestly, they will send the content of their  $val_i$ and $ts_i$ variable without changing them.
As a consequence, the reader will store locally in its $replies$ set the same pair $<v', ts>$ from any alive server.
Thus, at time $t_B(op)+2\delta \le t_B(op')+3\delta$, evaluating the condition at line \ref{read-C-06}, Fig. \ref{sfig:cread} the reader will select and return $v'$ and we have a contradiction.

\item {\bf Case 2.2 -  $\mathbf{t_B(op) <  t_B(op') + \delta}$.}\\
In this case, the {\sc write}$()$ message is concurrent (wrt. the happened before relation) with the {\sc read}$()$ message.
As a consequence, we may have that a server $s_i$ delivering the {\sc read}$()$ message before than the {\sc write}$()$ message and a server $s_j$ delivering the {\sc write}$()$ message before than the {\sc read}$()$ message.
As a consequence, $s_i$ will answer by sending a ${\sf reply}()$ message containing the old value $v\_old$ and the previous value (no more valid) while $s_j$ will answer by sending $v\_old$ and $v'$.
However, when the client evaluates the $replies$ set, it will find an occurrence of $v\_old$ for any alive server and evaluating the condition at line \ref{read-C-10}, Fig. \ref{sfig:cread} the reader will select and return $v\_old$ and we have a contradiction.

\item{\bf Case 2.3 -  $\mathbf{t_B(op')+2 \delta < t_B(op) < t_E(op')}$.}\\
In this case, the {\sc read}$()$ message may be delivered also after the end of $op'$. If no more ${\sf write}()$ operations occur before the end of $op$, we fall down to case 2.1 and the claim follows.
Contrarily, we fall down the the situation of case 2.2 where the concurrent write is a new write and the claim follow again.

\end{itemize}

\renewcommand{\toto}{lem:validityCorr}
\end{proofL}

\begin{theorem}\label{t:corretti}
Let $c_{\ell}$ be a client invoking a ${\sf read}()$ operation $op$. If all the servers behave honestly, the protocol shown in Fig. \ref{fig:readProtocol} - \ref{fig:detection} implements a regular register.
\end{theorem}

\begin{proofT}
The proof directly follows by Theorem \ref{th:termination} and Lemma \ref{lem:validityCorr}.
\renewcommand{\toto}{t:corretti}
\end{proofT}

\begin{lemma}\label{lemm:d2}
Let $c_j$ be a client sending a {\sc read}$()$ message at time $t$ and let {\sc reply}$(<i, ts, v, ots, ov>)$ be the message delivered by $c_j$ at time $t' >t$ as reply to the {\sc read}$()$.
Let $lts$ be the value stored locally in the $last\_ts$ variable by $c_j$ at time $t'$. 
If $| lts-ts | > 1$ then $s_i$ is malicious.
\end{lemma}

\begin{proofL}
Let us suppose by contradiction that $s_i$ sends to $c_j$ a {\sc reply}$(<i, ts, v, ots, ov>)$ such that $| lts - ts| > 1$ and $s_i$ is honest.

The $last\_ts$ local variable is updated to a certain value $x$ by any client $c_j$ when (i) $c_j$ delivered an {\sc ack}$(<j, x, ->)$ message from any server it believes honest and (ii) $x$ is strictly greater than the previous value stored in $last\_ts$ (line \ref{write-C-17}, Fig. \ref{sfig:cwrite}).

Let $t'$ be the time at which $c_j$ delivers the {\sc reply}$(<i, ts, v, ots, ov>)$ message and let us denote with $t_{up}$ the time at which $c_j$ updated its $last\_ts$ local variable to $lts$.
If $t_{ack}$ is the time at which $s_i$ delivered such {\sc write}$(v, lts)$ message and it sent back the {\sc ack}$(<i, lts, ->)$ message to clients, it follows that $t_{ack} < t_{up}$.

If at time $t'$ $lts \le ts$, it follows that $t_{up} \le t'$, otherwise $t_{up} > t'$. Let us consider the two cases separately.

\begin{itemize}
\item {\bf Case 1 - $\mathbf{t_{up} \le t'}$}. 
In this case $t' > t_{ack}$. 
Let us note that the value $ts$ sent by $s_i$ in the {\sc reply} message is the value stored locally by $s_i$ in the variable $ts_i$. 
Let us call $t_{rp}$ the time at which $s_i$ sends the {\sc reply}$(<i, ts, v, ots, ov>)$ message to $c_j$ and let us consider the following cases:

\begin{itemize}
\item {\bf Case 1.1 - ${\mathbf t_{rp} < t_{ack}}$}. In this case, $ts < lts$ and in particular, considering that ${\sf write}()$ operations are sequential, due to Corollary \ref{cor:singWriterNoCrash}, $ts=lts -1$. 
Thus, $| lts - ts| =1$ and we have a contradiction.

\item {\bf Case 1.2 - ${\mathbf t_{rp} > t_{ack}}$}. In this case, $ts = lts$ and again we have a contradiction.\\
\end{itemize}

\item {\bf Case 2 - $\mathbf{t_{up} > t'}$}. Let us note that the value $ts$ sent by $s_i$ in the {\sc reply} message is the value stored locally by $s_i$ in the variable $ts_i$. 
Let us call $t_{rp}$ the time at which $s_i$ sends the {\sc reply}$(<i, ts, v, ots, ov>)$ message to $c_j$ and let us consider the following cases:

\begin{itemize}
\item {\bf Case 2.1 - ${\mathbf t_{rp} < t_{ack}}$}. In this case, $ts = lts$ and we have a contradiction.

\item {\bf Case 2.2 - ${\mathbf t_{rp} > t_{ack}}$}. In this case, $ts < lts$ and in particular, considering that ${\sf write}()$ operations are sequential, due to Corollary \ref{cor:singWriterNoCrash}, $ts=lts -1$. 
Thus, $| lts - ts| =1$ and we have a contradiction.
\end{itemize}

\end{itemize}  

\renewcommand{\toto}{lemm:d2}
\end{proofL}

\begin{lemma}\label{lemm:d3}
Let $c_j$ be a client sending a {\sc read}$()$ message at time $t$ and let {\sc reply}$(<i, ts, v, ots, ov>)$ be the message delivered to $c_j$ at time $t' >t$ as reply to the {\sc read}$()$.
Let $mlts$ be the value stored locally in the $my\_last\_ts$ variable and let $mv$ be the value stored locally in the $my\_last\_value$ variable by $c_j$ at time $t'$. 
If $mlts=ts$ and $mv \neq v$ then $s_i$ is malicious.

\end{lemma}

\begin{proofL}
The proof simply follows by considering that a client $c_j$ updates its $my\_last\_ts$ and $my\_last\_val$ local variables only at the beginning of a ${\sf write}()$ operation.

Due to Corollary \ref{cor:singWriterNoCrash}, every ${\sf write}()$ operation has a unique timestamp thus if $c_j$ delivers a value $v \neq mv$ it means that the server altered it and it cannot be honest.

\renewcommand{\toto}{lemm:d3}
\end{proofL}

\begin{lemma}\label{l:valoreAtteso}
Let $c_j$ be a client sending a {\sc write}$(<j, ts, val>)$ message at time $t$. If there exists a server $s_j$ such that $s_j$ does not send the triple $<j,ts,val>$ by time $t+3\delta$, then $s_j$ is faulty.
\end{lemma}

\begin{proofL}
Considering that (i) the first {\sc read}$()$ request is sent at time $t+\delta$ (line \ref{write-C-06}, Fig. \ref{sfig:cwrite}), (ii) the ${\sf detection}()$ procedure on $replies$ set is performed at time $t+2\delta$ (line \ref{write-C-11}, Fig. \ref{sfig:cwrite}) and (iii) there not exists concurrent ${\sc write}()$ operations, then any honest server will always provide the expected triple and the claim follows.
\renewcommand{\toto}{l:valoreAtteso}
\end{proofL}

\begin{lemma}\label{l:ackAtteso}
Let $c_j$ a client sending a {\sc write}$(j, ts, val)$ message at time $t$. If there exists a server $s_j$ such that the triple $<j,ts,->$ does not appear in the $ack$ set at time $t+2\delta$, then $s_j$ is detected as faulty.
\end{lemma}

\begin{proofL}
Considering that (i) the ${\sf detection}()$ procedure on $ack$ set is performed at time $t+2\delta$ (line \ref{write-C-09}, Fig. \ref{sfig:cwrite}) and (ii) there not exists concurrent  ${\sf write}()$ operations, then a honest server will provide the expected triple and the claim follows.
\renewcommand{\toto}{l:ackAtteso}
\end{proofL}

\begin{theorem}\label{t:detection}
A honest alive server $s_i$ is never detected as malicious.
\end{theorem}

\begin{proofT}
A server $s_i$ deviates from the protocol if:
\begin{enumerate}
\item $s_i$ omits to send {\sc ack}$()$ messages during a ${\sf write}()$ or to sends {\sc reply}$()$ messages during a ${\sf read}()$ (lines \ref{detection-02}-\ref{detection-07}, Fig. \ref{fig:detection}).
\item $s_i$ sends bad timestamps (i.e. too old, lines \ref{detection-19}-\ref{detection-23}, Fig. \ref{fig:detection} or too new, lines \ref{detection-32}-\ref{detection-36}, Fig. \ref{fig:detection}).
\item $s_i$ sends a pair $<value, ts>$ with the correct timestamp and the wrong value  (lines \ref{detection-24}-\ref{detection-31}, Fig. \ref{fig:detection}).
\end{enumerate}

Thus, a honest server may be erroneously detected as faulty only if one of the previous cases occur.
However, due to Lemmas \ref{lemm:d2} - \ref{l:ackAtteso}, we have that if one of the previous situation happens, $s_i$ is necessarily malicious and the claim follows.


\renewcommand{\toto}{t:detection}
\end{proofT}

\begin{theorem}\label{th:final}
If $D_s, G_s$ and $\theta$ are such that equation \eqref{eq:AvsNA} holds then the protocol shown in Fig. \ref{fig:readProtocol} - \ref{fig:detection} implements a regular register with only $1$ alive honest server. 
\end{theorem}

\begin{proofT}
Let us first notice that the existence of the detection function creates the dualism between risky requests and risk-less request.
In addition, the existence of one honest alive server prevents the attacker to create a collusion able to make the condition at lines \ref{read-C-06}, \ref{read-C-10} and \ref{read-C15} satisfied with not valid values.
Thus, the claim simply follows by Lemma \ref{lem:GminD} and Theorem \ref{t:corretti}.
\renewcommand{\toto}{th:final}
\end{proofT}

\section{$\mathcal{P}_{cv}$ and $\mathcal{P}_{hash}$ Protocols for a Regular Register when $D_s \ge G_s$}\label{sec:other_protocols}

In the following, we show how to modify the protocol to get $\theta \ge \frac{1}{2}$, when $D_s \ge G_s$. In particular, we propose two possible extensions: the first using a probabilistic collaborative detection at the client side (introducing a cost in terms of number of messages needed to run the detection) and the second using a kind of fingerprint to prevent servers misbehavior (introducing a computational cost).

\noindent {\bf A collaborative detection protocol $\mathcal{P}_{cv}$.} The collaborative detection involves all the clients in the detection process and exploits the fact that the last writer remains in the system and it is always able to identify a faulty server. 
The basic idea is to look for a write witness (i.e., the writer) each time that a reader is not able to decide about the correctness of a value. This solution allows to identify malicious server and to decide and return always a correct value. 
However, considering that (i) we want to decouple as much as possible servers and client, (ii) this collaborative approach has a cost in terms of messages and (iii) to force rational servers to behave correctly it is sufficient to get $\theta \ge \frac{1}{2}$ (according to Lemma \ref{lem:GminD}), then we use this collaborative approach only with a given probability.
More in details, in $\mathcal{P}_{cv}$ protocol, when a reader does not collect the same value from all servers it flips a coin to decide if running the collaborative detection or not.
If the outcome is $1$, then it broadcasts to all the other clients the timestamps collected during the read operation and waits that some writer acknowledge them.
When a client receives a check timestamp request, it checks if it corresponds to its last written value and if so, it replies with such a value so that the reader can double-check information provided by servers.
If there is no match between values and timestamps, then clients are able to detect a faulty server and exclude it from the computation.\\
The introduction of this probabilistic step in the protocol increases the value of $\theta$ to $\frac{1}{2}$. As a consequence, following Lemma \ref{lem:GminD}, any rational server will decide to behave correctly to avoid to be detected.\\

\noindent {\bf A fingerprint-based detection protocol $\mathcal{P}_{hash}$.} Let us recall that the basic idea behind the detection process is to include inside reply messages (i.e., write acknowledgements or read replies) \vir{enough} information to verify the correctness of the provided information. In particular, in protocol $\mathcal{P}$, servers are required to acknowledge write operations by sending back the corresponding timestamp so that each client is always aware about it and the writer is able to verify that no bad timestamps are sent to clients.

In  protocol $\mathcal{P}_{hash}$, the basic idea is to extend $\mathcal{P}$ by including another information i.e., a fingerprint of the value and its timestamp (e.g., its hash), in the write message and in its acknowledgement so that it is always possible for a client to check that servers are replying correctly.
More in details, when a client writes, it computes the hash of the value and its corresponding timestamp and attaches such fingerprint to the message. In such way (as for $\mathcal{P}$) when servers acknowledge a write, they send back the correct fingerprint to all clients. 
Having such information, all clients are potentially able to detect locally if values collected during a read operation are never written values (this can be simply done by computing the hash of the message and compare it with the one received during the last write).
However, as in the case of $\mathcal{P}_{cv}$, this detection has a cost and, to get $\theta \ge \frac{1}{2}$ it is sufficient that this is done with a certain probability.
Thus, when a reader does not collect the same value from all servers, it flips a coin and if the outcome is $1$ then it computes the hash of the messages it delivered and compares them with the hashes it knows to be associated to a specific timestamp. The introduction of this step is enough to get $\theta = \frac{1}{2}$ and to prevent rational servers deviating from the protocol. Notice that, as for $\mathcal{P}_{cv}$, the employment of the random coin has a twofold purpose: (i) to provide a solution for $D_s \ge G_s$, for which it is enough to have $\theta \ge \frac{1}{2}$ and (ii) to avoid to always perform the costly detection operation.\\

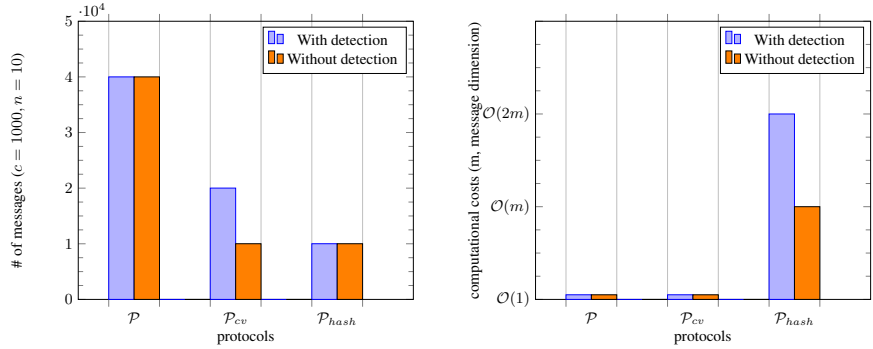
\begin{figure}[t]

\begin{tikzpicture}[scale=.65]
\begin{axis}[ybar interval, ymax=50000,ymin=0, minor y tick num = 3, xmax=30, bar shift=0pt, ylabel={\# of messages ($c=1000,n=10$)},  xticklabels={,,}, xlabel=protocols, xticklabels={$\mathcal{P}$, ,$\mathcal{P}_{cv}$,,$\mathcal{P}_{hash}$}]
\addplot coordinates { (0, 40000) (5, 0) (10, 20010) (15, 0) (20,10010) (25,0) };
\addlegendentry{With detection}
\addplot[fill=orange] coordinates { (0, 40000) (5, 0) (10, 10010) (15, 0) (20,10010) (25,0) };
\addlegendentry{Without detection}
\end{axis}
\end{tikzpicture} %
\hspace{.5cm}
\begin{tikzpicture}[scale=.65]
\begin{axis}[ybar interval, ymax=6,ymin=0, minor y tick num = 3, xmax=30, bar shift=0pt, ylabel={computational costs (m, message dimension)},  xticklabels={,,}, xlabel=protocols, xticklabels={$\mathcal{P}$, ,$\mathcal{P}_{cv}$,,$\mathcal{P}_{hash}$}, yticklabels={,$\mathcal{O}(1)$,$\mathcal{O}(m)$,$\mathcal{O}(2m)$}]
\addplot coordinates { (0, .1) (5, 0) (10, .1) (15, 0) (20,4) (25,0) };
\addlegendentry{With detection}
\addplot[fill=orange] coordinates { (0, .1) (5, 0) (10, .1) (15, 0) (20,2) (25,0) };
\addlegendentry{Without detection}
\end{axis}
\end{tikzpicture}
\caption{Qualitative analysis of protocols with respect to their message complexity (left figure) and computational complexity (right figure).
For the message complexity we consider a system where the number of servers is $n=10$ and the number of clients is $c=1000$. 
For the computational complexity we consider the cost with respect to the message size $m$.}\label{fig:tradeoff}
\end{figure}

\noindent{\bf Trade offs.}
Figure \ref{fig:tradeoff} shows a qualitative comparison of the three proposed protocols in terms of message complexity and computational cost. In particular, we compare the cost of the protocols both in presence and absence of a server attack (i.e., when the detection is necessary or not).
As we can see, $\mathcal{P}$ requires the highest number of messages and such number does not depend on the real need of doing detection but it is rather required to mask the type of operation that a client is doing and to make indistinguishable real read messages from dummy ones. Concerning 
its computational cost, it is constant since it does not depend on the message size.\\ 
In $\mathcal{P}_{cv}$ it is possible to save the dummy read messages as we do not need anymore to mask the message pattern but we need to pay the cost of the collaborative detection, if it is needed. In fact, if a reader is not able to decide a value, it needs to send messages to contact all the other clients (higher message complexity in case of server misbehaviour). Concerning the computational cost, it is not affected by the detection.
Conversely, $\mathcal{P}_{hash}$ exhibits the dual behaviour: message complexity is not affected by server misbehaviour but the computational cost is impacted by the need of detection.

Thus, we can conclude saying that $\mathcal{P}$ is a pessimistic protocol and is the most expensive one but it allows to maintains clients and servers completely decoupled. Contrarily, $\mathcal{P}_{cv}$ and $\mathcal{P}_{hash}$ are optimistic as they perform lightweight operations and, if needed, they perform an heavy detection (with a high message cost in the case of $\mathcal{P}_{cv}$ and a high computational cost in case of $\mathcal{P}_{hash}$).

\subsubsection{Sketch of $\mathcal{P}_{cv}$ and$\mathcal{P}_{hash}$ correctness\\}%
\noindent%
$\mathcal{P}_{cv}$ has not variations with respect to $\mathcal{P}$ concerning the write and read operations (the detection procedure is removed from the write operation). Thus the algorithm is proved to work. Concerning the detection part we have that when a clients fails to read, then it asks to other clients to reach the last writing one that replies with the correct value. Thus the reading clients is able to detected servers acting in a malicious way. Since the detection procedure is run depending on the outcome of a random bit then $\theta=\frac{1}{2}$. Thus if $D_s>G_s$ then \eqref{eq:AvsNA} holds then $\mathcal{P}_{cv}$ implements a regular register with only $1$ alive honest server.\\
\\
$\mathcal{P}_{hash}$ has not variations with respect to $\mathcal{P}$ concerning the write and read operations (the detection procedure is removed from the write operation). Thus the algorithm is proved to work. The difference with  $\mathcal{P}$ is that the detection procedure slightly modifies the write operation. Each time a new value is written, the hash is computed and spread along with the timestamp. The proof of that directly follows generalizing Lemma \ref{lem:ackDetection}. 
When a clients fails to read then it performs the hash of the delivered replies to compare them to the one it knows to be associated to the related timestamp. Thus the reading clients is able to detected servers acting in a malicious way. Since the detection procedure is run depending on the outcome of a random bit then $\theta=\frac{1}{2}$, if $D_s>G_s$ then \eqref{eq:AvsNA} holds then $\mathcal{P}_{cv}$ implements a regular register with only $1$ alive honest server.\\

\section{Conclusion}\label{sec:conclusion}
This paper addresses the problem of building a regular register in a distributed system where clients are anonymous and servers maintaining the register state may be rational malicious processes.
We have modelled our problem as a two-parties Bayesian game and we designed distributed protocols able to reach the Bayesian Nash Equilibrium and to emulate a regular register when the loss in case of detection is greater than the gain obtained from the deviation (i.e. $D_s > G_s$).
To the best of our knowledge, our protocols are the first register protocols working in the absence of knowledge on the number of compromised replicas.

The protocols rely on the following assumptions: (i) rational malicious servers act independently and do not form a coalition, (ii) the system is synchronous, (iii) clients are anonymous and (iv) write operations are serialised. 

As future works, we are investigating how to solve the same problem under weaker synchrony assumption or in the case an attacker controls a coalition of processes. Addressing these points is actually far from be trivial.
Considering a fully asynchronous system, in fact, makes impossible to use our punishment mechanism as clients are not able to distinguish alive but silent servers from those crashed. 
Additionally, when the attacker is able to compromise and control a coalition of processes, the model provided in this paper is no more adeguate and we are studying if and how it is possible to define a \emph{Bayesian Coalitional Game} \cite{IS08} for our problem and if an equilibrium can be reached in this case.

\section*{Acknowledgments}
This present work has been partially supported by the EURASIA project and CINI Cybersecurity National Laboratory within the project FilieraSicura: Securing the Supply Chain of Domestic Critical Infrastructures from Cyber Attacks (www.filierasicura.it) funded by CISCO Systems Inc. and Leonardo SpA.


%


\newpage

\begin{thebibliography}{99}
%
\bibitem{ADH13}
Abraham, I.,  Dolev, D., Halpern, J. Y. 
{\it Distributed Protocols for Leader Election: A Game-Theoretic Perspective.} 
DISC 2013: 61-75

\bibitem{AGLM14}
Afek, Y., Ginzberg, Y., Landau Feibish, S. and Sulamy, M. 
{\it Distributed computing building blocks for rational agents.} 
PODC 2014: 406-415.

\bibitem{AAC05}
Aiyer, A. S., Alvisi, L., Clement, A., Dahlin, M., Martin, J. P., and Porth, C.
{\it BAR fault tolerance for cooperative services}. 
ACM SIGOPS Operating Systems Review. ACM, 2005. p. 45-58.

\bibitem{ABD95}
Attiya, H., Bar-Noy, A., and Dolev, D. 
{\it Sharing memory robustly in message-passing systems}. 
Journal of the ACM 42, 1, 1995, 124-142.

\bibitem{B00}
Bazzi R. A., 
{\it Synchronous Byzantine Quorum Systems}, 
Distributed Computing 13(1), 45-52, 2000.

\bibitem{CKW00}
Chaudhuri S., Kosa M.J. and Welch J., 
{\it One-write Algorithms for Multivalued Regular and Atomic Registers.} 
Acta Informatica, 37:161-192, 2000. 

\bibitem{CLNMAD08}
Clement, A., Li, H. C., Napper, J., Martin, J., Alvisi, L., Dahlin, M. 
{\it BAR primer},
DSN 2008: 287-296

\bibitem{CNLMAD07}
Clement, A., Napper, J., Li, H., Martin, J. P., Alvisi, L., and Dahlin, M.
{\it Theory of BAR games},
PODC 2007: 358-359


\bibitem{DFT13}
Delporte-Gallet, C., Fauconnier, H., Tran-The, H.
{\it Uniform Consensus with Homonyms and Omission Failures}
ICDCN 2013: 161-175

\bibitem{FT91}
Fudenberg, D., Tirole, J. 
{\it  Game theory}, 1991. Cambridge, Massachusetts.

\bibitem{HV95}
Haldar S. and Vidyasankar K., 
{\it Constructing 1-writer Multireader Multivalued  Atomic Variables
from Regular Variables. }
JACM, 42(1):186-203, 1995. 

\bibitem{IS08}
Ieong, S., Shoham, Y.
{\it Bayesian Coalitional Games.} 
AAAI. 2008:  95-100

\bibitem{LCWNRAD06}
Li, H. C., Clement, A., Wong, E. L., Napper, J., Roy, I., Alvisi, L., Dahlin, M.
{\it BAR Gossip} 
OSDI 2006: 191-204

\bibitem{L86}
Lamport. L.,
On Interprocess Communication,  Part 1: Models, Part 2: Algorirhms,
{\em Distributed Computing}, 1(2):77-101, 1986.

\bibitem{MSLCADW11}
Mahajan, P., Setty, S., Lee, S., Clement, A., Alvisi, L., Dahlin, M., and Walfish, M.
 {\it Depot: Cloud storage with minimal trust}, 
ACM TOCS 29(4), 2011

\bibitem{MR98}
Malkhi D., Reiter M. K. 
{\it Byzantine Quorum Systems}, 
Distributed Computing 11(4), 203-213, 1998.

\bibitem{MAD02}
Martin J., Alvisi L., Dahlin M.,
{\it Small Byzantine Quorum Systems},
DSN 2002: 374-388.

\bibitem{MAD02-2}
Martin J., Alvisi L., Dahlin M.. 
{\it Minimal Byzantine Storage}, 
DISC 2002.

\bibitem{S90}
Schneider, F. B. 
{\it Implementing fault-tolerant services using the state machine approach: A tutorial},
ACM Computing Surveys 22(4): 299-319, 1990.

\bibitem{SAG94}
Singh A.K., Anderson J.H. and Gouda M., 
{\it The Elusive Atomic Register. }
JACM, 41(2):331-334, 1994. 

\bibitem{SBCNV10}
Sousa, P., Bessani, A. N., Correia, M., Neves, N. F., Verissimo, P.
{\it Highly available intrusion-tolerant services with proactive-reactive recovery}, 
IEEE TPDS 21(4): 452-465, 2010 .

\bibitem{TOR}
The Tor Project \url{https://www.torproject.org}.

\bibitem{V88}  
Vidyasankar K., 
{\it Converting Lamport's Regular Register to Atomic Register.} 	
IPL, 28(6):287-290, 1988

\bibitem{VA86} 
Vityani P. and Awerbuch B., 
{\it Atomic Shared Register Access by Asynchronous Hardware. }
FOCS 1987, 223-243.

\bibitem{OY91}
Ostrovsky, R. and Yung, M.,
{\it How to withstand mobile virus attacks.}
PODC 1991, 51-59.

\bibitem{S79}
Shamir, A.
{\it How to share a secret}
Comm. of ACM 1979, 612-613.

\bibitem{DEL16} 
Dolev, S., ElDefrawy, K., Lampkins, J., Ostrovsky, R., and Yung, M.
{\it Proactive secret sharing with a dishonest majority. }
SCN 2016, 529-548.

\bibitem{CD15} 
Cramer, R. and Damg\:{a}rd, I. B.
{\it Secure Multiparty Computation}
Cambridge University Press 2015.




\end{thebibliography}
\end{document}